\documentclass[aps,pra,reprint,11pt,superscriptaddress,notitlepage,nofootinbib,onecolumn,showkeys,longbibliography]{revtex4-1}

\usepackage[utf8]{inputenc}
\usepackage[T1]{fontenc}
\usepackage{lmodern}
\usepackage{microtype}

\usepackage{amsmath}
\usepackage{graphicx}
\usepackage{url}
\usepackage{hyperref}
\hypersetup{
  colorlinks=true,
  citecolor=blue,
  linkcolor=blue,
  urlcolor=blue,
  breaklinks=true
}

\usepackage{comment}

\newcommand\snowmass{
  \begin{center}
    \rule[-0.2in]{\hsize}{0.01in}\\\rule{\hsize}{0.01in} \\
    \vskip 0.1in
    Submitted to the  Proceedings of the US Community Study \\
    on the Future of Particle Physics (Snowmass 2021) \\ 
\rule{\hsize}{0.01in} \\
\rule[+0.2in]{\hsize}{0.01in} \end{center}
}

\begin{document}
\snowmass


\title{Detector and Beamline Simulation for Next-Generation High Energy Physics Experiments}

\author{Sunanda Banerjee}
\affiliation{University of Wisconsin, Madison, WI 53706, USA}
\author{D. N. Brown}
\affiliation{Lawrence Berkeley National Laboratory, Berkeley, CA 94720, USA}
\author{David N. Brown}
\affiliation{Western Kentucky University, Bowling Green, KY 42101, USA}
\author{Paolo Calafiura}
\affiliation{Lawrence Berkeley National Laboratory, Berkeley, CA 94720, USA}
\author{Jacob Calcutt}
\affiliation{Oregon State University, Corvallis, OR 97331, USA}
\author{Philippe Canal}
\affiliation{Fermi National Accelerator Laboratory, Batavia, IL 60510, USA}
\author{Miriam Diamond}
\affiliation{University of Toronto, Toronto, Ontario M5S 1A7, Canada}
\author{Daniel Elvira}
\affiliation{Fermi National Accelerator Laboratory, Batavia, IL 60510, USA}
\author{Thomas Evans}
\affiliation{Oak Ridge National Laboratory, Oak Ridge, TN, 37831, USA}
\author{Renee Fatemi}
\affiliation{University of Kentucky, Lexington, KY 40506, USA}
\author{Krzysztof Genser}
\email{genser[AT]fnal.gov}
\affiliation{Fermi National Accelerator Laboratory, Batavia, IL 60510, USA}
\author{Robert Hatcher}
\affiliation{Fermi National Accelerator Laboratory, Batavia, IL 60510, USA}
\author{Alexander Himmel}
\affiliation{Fermi National Accelerator Laboratory, Batavia, IL 60510, USA}
\author{Seth R.~Johnson}
\affiliation{Oak Ridge National Laboratory, Oak Ridge, TN, 37831, USA}
\author{Soon Yung Jun}
\affiliation{Fermi National Accelerator Laboratory, Batavia, IL 60510, USA}
\author{Michael Kelsey}
\affiliation{Texas A\&M University, College Station, TX 77843, USA}
\author{Evangelos Kourlitis}
\affiliation{Argonne National Laboratory, Lemont, IL 60439, USA}
\author{Robert K. Kutschke}
\affiliation{Fermi National Accelerator Laboratory, Batavia, IL 60510, USA}
\author{Guilherme Lima}
\affiliation{Fermi National Accelerator Laboratory, Batavia, IL 60510, USA}
\author{Kevin Lynch}
\affiliation{York College and the Graduate School, The City University of New York, New York, NY 11451, USA}
\author{Kendall Mahn}
\affiliation{Michigan State University, East Lansing, MI 48824, USA}
\author{Zachary Marshall}
\affiliation{Lawrence Berkeley National Laboratory, Berkeley, CA 94720, USA}
\author{Michael Mooney}
\affiliation{Colorado State University, Fort Collins, CO 80523, USA}
\author{Adam Para}
\affiliation{Fermi National Accelerator Laboratory, Batavia, IL 60510, USA}
\author{Vincent R. Pascuzzi}
\email{pascuzzi[AT]bnl.gov}
\affiliation{Brookhaven National Laboratory, Upton, NY 11973, USA}
\author{Kevin Pedro}
\affiliation{Fermi National Accelerator Laboratory, Batavia, IL 60510, USA}
\author{Oleg Samoylov}
\affiliation{Joint Institute for Nuclear Research, Dubna, 141980, Russia}
\author{Erica Snider}
\affiliation{Fermi National Accelerator Laboratory, Batavia, IL 60510, USA}
\author{Pavel Snopok}
\affiliation{Illinois Institute of Technology, Chicago, IL 60616, USA}
\author{Matthew Szydagis}
\affiliation{University at Albany SUNY, Albany, NY 12222, USA}
\author{Hans Wenzel}
\affiliation{Fermi National Accelerator Laboratory, Batavia, IL 60510, USA}
\author{Leigh H. Whitehead}
\affiliation{University of Cambridge, Cambridge, CB3 0HE, United Kingdom}
\author{Tingjun Yang}
\affiliation{Fermi National Accelerator Laboratory, Batavia, IL 60510, USA}
\author{Julia Yarba}
\affiliation{Fermi National Accelerator Laboratory, Batavia, IL 60510, USA}

\begin{abstract}
The success of high energy physics programs relies heavily on accurate detector simulations and beam interaction modeling.
The increasingly complex detector geometries and beam dynamics require sophisticated techniques in order to meet the
demands of current and future experiments.
Common software tools used today are unable to fully utilize modern computational resources, while
data-recording rates are often orders of magnitude larger than
what can be produced via simulation.
In this paper, we describe the state, current and future needs of high energy physics detector and beamline simulations and related challenges, and we propose a number of possible ways to address them.
\end{abstract}

\date{March 31, 2022}
\maketitle

\section{Introduction}
\label{sec:intro}

Detector simulations, modeling of beam interactions and particle production in fixed targets
are indispensable in the design process of new detectors and facilities for High Energy Physics
(HEP) experiments.  Equally important is the role these simulations play in the development
of reconstruction algorithms, and validation and interpretation of experimental results.
Challenges related to both underlying physics and practical computation
arise from the increasing complexity of HEP detectors and experiments,
which are being designed to be more precise and to detect and measure rarer processes.
For example, searches for beyond the Standard Model physics may require implementing new models or refining current ones, while technical challenges also emerge as a
result of the growing number of heterogeneous high performance computing (HPC)
platforms being built and supported under the purview of the DOE.

\section{Simulation needs and their drivers}
\label{sec:needs}

\subsection{Computing Hardware Evolution}
\label{subsec:hardwareevolution}

The changing computing hardware landscape imposes new constraints on the way the simulations
can and need to be performed.  These new constraints make the old computing processing
model---where separate instances of often general-purpose simulation applications were run on their own computer cores, often called ``embarrassingly parallel''---more
difficult, if not impossible, to sustain. As the market-driven trends in hardware---which are not
significantly influenced by HEP---continue to evolve, the simulation software, algorithms, and
techniques must also evolve in order to allow the use of the available computing infrastructure:
in particular, new programming models, compilers, and software libraries.
Research and development (R\&D) into
more accurate models, and more efficient and versatile codes, and efficient use of modern
hardware---e.g., GPUs, FPGAs, and other hardware accelerators---is paramount to the continued
success of HEP experiments.

\subsection{Evolving Physics-Related Needs}
\label{subsec:physicsneeds}

The complexity of detectors and sizes of datasets increase, physics measurements
and theoretical predictions can be made with higher precision, and experiments seek sensitivity
to rarer processes. Thus, there is a rapid growth in demand for larger numbers of simulated events with higher fidelity. While each of
the experimental HEP frontiers has somewhat different
requirements for improvements in detector simulations, there are a
number of overlapping needs. For some experiments, increases in detector complexity---for
instance, to accommodate higher beam energy, luminosity, beam intensity conditions, and per-event
multiplicities---will drive the development of parameterized or machine-learned models,
and other novel techniques for speed-ups of detector simulation. For other experiments, the demand
for more accurate simulations---such as needed for high-fidelity modeling of signal induction in
liquid Argon (LAr) \cite{LEPLAR} and other scintillation-based materials, Cerenkov light
propagation, condensed matter effects, low-energy response, and rare background processes--- 
will require the development of additional and more complete physics models.
Apart from detector
simulations, some experiments depend critically upon a detailed and precise understanding of
the complex interactions of beams with fixed targets, such as needed for secondary beam
production, stopping particles, etc.

\section{Current simulation tools and the central role played by Geant4}
\label{sec:currenttools}

While HEP experiments use variety of tools to perform detector simulations,
Geant4~\cite{Agostinelli:2002hh,Allison:2006ve,Allison:2016lfl} is a toolkit used by most, if
not all, of them for at least some elements of those simulations. Over more than the last twenty years,
Geant4, building upon the experience gained with GEANT~3~\cite{GEANT3}, has become a de-facto standard
for many aspects of the HEP detector simulations.

Geant4 may be augmented by other
physics packages, such as NEST~\cite{NEST}, G4CMP~\cite{G4CMP}, Opticks~\cite{Simon_C_2017} or geometry
related packages---e.g., VecGeom~\cite{VecGeom}, CADMesh~\cite{CADMesh}---with GDML~\cite{GDML}, usually used 
to exchange geometry and material information among software components (with
DD4hep~\cite{DD4hep} starting to be used as well). Other packages, such as FLUKA~\cite{FLUKA:2014},
MARS~\cite{MARS:2007}, and MCNP~\cite{MCNP62}, are also employed e.g., to crosscheck Geant4 results or to
make specialized calculations.

While Geant4 fulfills most of the current needs, and is supported by a worldwide collaboration, it
is being pushed to its limits by new experiments and changing computing hardware.
To address missing Geant4 features, many experiments extend or replace Geant4 models with specialized code.
Most such developments are needed for experiments outside of the HEP Energy Frontier for which
Geant4 had predominantly been developed initially.

\subsection{Extending Geant4 by interfacing other packages with it}
\label{subsec:crossDependencies}

To enable more accurate detector simulations, additional packages can be used to
enhance Geant4 capabilities. The list of such packages, augmenting Geant4, includes: NEST,
G4CMP, Opticks, and PYTHIA 8~\cite{PYTHIA8:2014}. 
This is in addition to event generator packages that are used to define primary particles for a given event.
A very brief description of each package and how it can be used with Geant4 follows.

\begin{itemize}

\item{NEST} -- the Noble Element Simulation Technique simulates  excitation, ionization, and other processes in noble elements, providing calculations of photon and electron yields and fluctuations with energy- and field-dependent empirical models. It can be used standalone or as an extension of Geant4.

\item{PYTHIA 8} -- a general purpose Monte Carlo event generator, that can be used to generate
  high-energy physics collision events as input to Geant4. It can also be
  interfaced with Geant4 to decay, e.g., charm, beauty, or tau particles, or to replace certain Geant4
  decay tables to assure consistency of the decays when the two packages are used together in
  an experimental framework.

\item{G4CMP} -- a Condensed Matter Physics for Geant4 package, which, among other features, models the
  production of electron-hole pairs and phonons from energy deposits and the subsequent
  transport and interactions of the produced objects.

\item{Opticks} -- a GPU Accelerated Optical Photon Simulation using NVIDIA OptiX GPU ray tracing
  library, can be used with Geant4 to enable fast simulation of optical photon generation and transport, 
  generate photon look-up tables, or eliminate the use of pre-generated photon look-up tables altogether.

\end{itemize}

In addition to performing simulations with the above packages, some experiments need to, for example,
simulate processes such as detailed transport of electrons or ions in gaseous or liquid media taking into 
account precise field calculations, which goes beyond the current capabilities of Geant4.

An example of a  slightly different package, used to make fast relative comparisons of certain detector configurations, which can be used after running Geant4 is
TrackToy~\cite{TrackToy:0.3.1} -- a hybrid Monte Carlo that takes as input particle 4-vectors from a Geant4 simulation,
and makes a simplified simulation of detector geometry, material, and Kalman filter track reconstruction. Another example is Geant4Reweight~\cite{Calcutt:2021zck} -- a framework for evaluating and propagating hadronic interaction uncertainties in Geant4.
 
\section{Elaboration on some specific detector simulation needs}
\label{sec:needs2}

In order to fully leverage the work which went into development of the above software packages,
one needs to constantly support them as well as related Geant4 interfaces enabling their use, 
to make sure the packages stay current and keep up with the ongoing simultaneous developments 
and new requirements. The people developing the packages and Geant4 need to be supported continuously as well (see Section~\ref{sec:experts}).

\subsection{Some needs related to hadronic interaction modeling}
\label{subsec:modelingNeeds}

HEP experiments in which the beam energies are such that a modeling of hadronic particle
interactions below 10~GeV is required rely predominantly on Geant4 models---such as Bertini-like
intranuclear cascade~\cite{Bertini:2015}, Fritiof (FTF)~\cite{FTF:1,FTF:2} string, Geant4 precompound, evaporation and breakup~\cite{Allison:2016lfl}---to simulate the relevant interactions. 
In the study~\cite{Geant4vmp} aiming to improve Geant4 physics models' agreement with data, and to
provide ways to estimate simulation systematic uncertainties by varying model
parameters, it has been determined that varying model parameters
can lead to substantially better agreement with some datasets.
However, more degrees of freedom
are required for better overall agreement, which means that some models need more work to describe the existing data.
Unfortunately, the Bertini model has not been actively developed over the last
few years due to the lack of personpower. This is especially alarming given that almost all
current HEP experiments rely on it. Therefore, one needs to resume and/or expand the work on a relatively short 
timescale before the gap in the model development starts to critically affect current and future experiments. 
In addition, new experiments, either reaching out to higher energies or searching for very rare processes, require
improvements in the range, precision, and flexibility of the current models, as well as implementation of the new ones.
These efforts would all benefit if additional personpower were available to contribute
(see more comments regarding the human challenge in Section~\ref{sec:experts} and \mbox{Ref.}~\cite{SmallHEPexp}, Section 2.4 on Support for Common Tools ).

\subsection{Dedicated measurements benefiting physics model developments}
\label{subsec:dedicatedMeasurements}
 
The Geant4 parameter studies mentioned in the previous
subsection~(\ref{subsec:modelingNeeds}) and related physics model
developments are impossible without the availability of experimental
data that can be used to validate the models.  One set of measurements of the negative pion total hadronic cross section on Argon (for the pion
kinetic energies in the range 100--700 MeV) was published by the
LArIAT experiment~\cite{LArIAT:2021yix} recently.
The ProtoDUNE experiment~\cite{DUNE:2020cqd, DUNE:2021hwx} plans to
publish results on the hadron-Argon cross sections for $\pi^{+}$,
proton, and $K^{+}$ in the momentum range 0.3--7 GeV/c, including
processes of elastic scattering, quasielastic scattering, and
various inelastic scattering channels (such as absorption and charge exchange), as well as the kinematic distributions of the final
state particles.
 
Such measurements using the Argon target are especially valuable
because the data can be used to improve the modeling of hadron-Argon
interactions. This will help to reduce the simulation systematic
uncertainties for the current and future neutrino experiments using liquid Argon time projection chambers, in particular the
short- and long-baseline neutrino experiments at Fermilab~\cite{Machado:2019oxb, DUNE:2020lwj} 
(MicroBooNE, SBND, ICARUS and DUNE).
These examples illustrate why measurements of basic quantities needed for validation and development of physics models
are very important. They should be considered an integral part of
the design process of new experiments, in order to be able to accurately
simulate them and interpret their results.

\section{Multi-prong Approach}
\label{sec:multiprong}

In order to satisfy the stringent and complex requirements of near-term and future programs, and
to meet the computing challenges within the resource budgets, the HEP detector simulation community
has undertaken a multi-prong R\&D and operations program (see, e.g., \mbox{Ref.~}\cite{Albrecht_2019}).
In the case of detector and beam simulations, elements of the program involve adaptation and
extensions of the existing software as well as efforts to write new packages or modules to
eliminate constraints resulting from many years of evolution of legacy codes.

\subsection{Specific R\&D Efforts}
\label{subsec:specificRD}

Some of the elements of a proposed R\&D program which play to the strengths of the US
simulation teams are listed below (and described in subsequent sections or their own, separate
white papers or other documents). More information on machine learning based projects is provided in Section~\ref{subsec:fastsimML} below.

\begin{itemize}
  
\item{``Celeritas'' --  a project that implements a growing set of physics for detector simulation targeted at GPU-powered HPC platforms~\cite{johnson_2021,Celeritas:SM}; a component of the project, "Acceleritas", provides interfaces between Geant4 and Celeritas, to enable a hybrid CPU/GPU workflow with selected tasks executed on the GPUs;}

\item{``Simulating Optical Photons in HEP experiments on GPUs'' -- is an effort to integrate
    recent versions of Geant4 and Opticks~\cite{Simon_C_2017,opticks-LOI} in a hybrid CPU/GPU
    application using Geant4 Tasking a Task-Level Parallelization approach introduced in Geant4
    v11~\cite{Geant4:11,G4Tasking-LOI,CaTS2021};}

\item{``Simulation on HPCs'' -- proposes to investigate the interconnection of HPC systems for
    event simulation and task scheduling~\cite{hpcsim-LOI};}

\item{``Simulations of Low-Energy Crystal Physics for Dark Matter Detectors'' -- describes the
    needs of a certain class of Dark Matter (DM) experiments and presents ideas on how to
    address them~\cite{LECPDM-LOI};}

\item{``Optimizing Geant4 parameters and enabling estimating related systematic
    uncertainties''} -- is an activity described in \mbox{Ref.~}\cite{Geant4vmp}.

As the projects need to be seen in a worldwide HEP context, one should also mention
two related European HEP efforts:

\item{``G4HepEm'' -- an R\&D project to make electron/positron/gamma
transport faster by restructuring, specializing and separating
underlying libraries, targeting optimization of execution on CPU as
well as on GPU~\cite{G4HepEm:1,G4HepEm:2}}

and 

\item{``AdePT'' (Accelerated demonstrator for electromagnetic Particle
Transport) -- an R\&D project to transport electrons/positrons/gammas
on GPUs; it makes use of G4HepEm and VecGeom, with the latter also being
redesigned to improve its GPU performance; AdePT~\cite{AdePT} is being integrated with
Geant4 to offload processing of electrons/positrons/gammas to GPUs}.

\end{itemize}

\subsection{Fast Simulations and Machine Learning}
\label{subsec:fastsimML}

The evolution of fast simulation tools and techniques is another element of the multi-pronged
approach. Machine learning (ML) is one of the most promising alternatives to traditional
parameterization-based methods.
In particular, ML algorithm inference can naturally be accelerated on coprocessors such as GPUs or FPGAs, providing an alternative method to utilize new computing hardware.
However, speedups obtained this way are only valid if the resulting output accurately reproduces the original physics of particle-detector interactions.
In addition, ML-based simulations still require computationally
expensive training campaigns that will necessarily rely heavily on some well-established simulation
software, most likely Geant4. Therefore, ML does not eliminate the need to establish long-term efforts
to improve the speed of existing software and to make it run efficiently on accelerators or HPC hardware.

An ML algorithm can enter the simulation workflow in different ways:
it can replace or augment part or all of Geant4, or part or all of a traditional fast simulation.
Relevant types of ML algorithms include generative adversarial networks (GANs), variational autoencoders (VAEs), normalizing flows (NFs), regression-based approaches, and other, more complicated and harder to classify methods.
Each approach and technique has different benefits and challenges, which are discussed in more detail in \mbox{Ref.~}\cite{ML4Sim_White_Paper} and related work.
In all cases, the reliability of the ML algorithm must be carefully assessed
to ensure that extrapolation beyond the training data is valid and that physically inaccurate events are not produced.
The first production-scale test of GANs as a fast simulation tool will be conducted by ATLAS during LHC Run 3~\cite{ATLAS:2021pzo} and will provide more insight into these questions.
While existing explorations of these techniques have focused primarily on collider physics,
they may also prove useful for other subfields of experimental HEP in the future.
Some of the advantages of ML, such as fast evaluation on coprocessors, as well as other benefits can be further exploited by employing differentiable programming techniques more widely in detector simulation software.
This relatively new option is also discussed further in \mbox{Ref.~}\cite{ML4Sim_White_Paper}.

\section{Need for experts and their training (the ``human challenge'')}
\label{sec:experts}

Detector simulation tools require effort for code modernization, improvement of physics models,
maintenance, and long-term support.  Evolving computing architectures demand significant
investment to adapt and optimize simulation software to run efficiently and exploit the
available hardware at modern HPC centers.  The above efforts are not viable without
comprehensive and intensive training plans for both application developers and end users.

Due to the long lifetimes of current and future HEP experiments, it is crucial to continuously
recruit, train, and retain teams of experts, as well as to create attractive career paths for
people developing and maintaining the software. Most importantly, detector modeling
requires multidisciplinary teams consisting of software developers and physicists with
specializations in low-energy, electromagnetic, and weak interactions, as well as condensed matter
physics. Continuous funding is therefore required for High-Energy and Nuclear physicists in these roles, as
well as software developers throughout the life cycles of HEP software toolkits.

To set the scale of the required detector simulation R\&D and operations effort, we can use the
Geant4 toolkit and the GeantV R\&D project~\cite{amadio2020geantv} as examples. Geant4 is
maintained by an almost 30-year-old collaboration of well more than 100 members (and more than
30 FTEs) distributed worldwide, serving a very diverse set of user domains. Its developer and
user bases extend from HEP to astronomical and radiation studies and medical applications. GeantV was
an R\&D project to redesign Geant4 to exploit the benefits of vectorization and increased code
and data locality. It took about 5 years (it ended in 2020) and 30 FTE-years to implement only a fraction of all
Geant4 modules within GeantV, mainly in the transport and electromagnetic domains, leaving most
of the code unvectorized (the cited paper describes the lessons learned from writing the prototype). Based on this recent R\&D experience and the personpower requirements implied by the size of the
international Geant4 collaboration, it is critical that the US HEP community contributes a substantial portion of the
global detector simulation effort, commensurate with the needs. A team of highly-skilled physicists and engineers is
required to provide the necessary support and developments for Geant4, and the packages extending it, 
to meet the needs and
challenges within the scope of the US HEP experimental program.
This must also include sufficient support for developers with domain expertise in the software frameworks of HEP experiments, who can properly and efficiently utilize new or updated simulation toolkits provided by the Geant4 collaboration and/or other detector simulation groups.
The realization of Geant4
running on exascale computing hardware will require additional personpower---a team whose
expertise lies in parallel and high-performance computing---to deliver a production-quality
framework on the timescale of the High-Luminosity Large Hadron Collider
program~\cite{Apollinari:2284929} and future Neutrino Physics, Rare Processes and 
Precision~\cite{Abi:2020wmh,CLFV, Mu2e2} and Cosmic~\cite{dm-new-inits} Frontier experiments.

\section{Summary}
\label{sec:summary}

We have described some of the current and future detector and beamline simulation needs (Section~\ref{sec:needs} and~\ref{sec:needs2}).
The main drivers of these needs are the simulation speed and accuracy, the ability to run and efficiently
use current and future computing hardware, and the ability to adequately simulate all important
physics processes. We mentioned dedicated measurements enabling validation of the physics
models and the need to perform them (Subsection~\ref{subsec:dedicatedMeasurements}).
We described the simulation software in use (Section~\ref{sec:currenttools}), with Geant4 and the packages
used with it playing a central role in the design of new detectors and facilities and in the
development of reconstruction algorithms, as well as in the validation and interpretation of 
experimental results.
We observed that the widespread use of Geant4 and the evolving landscape of computing hardware motivates the need to continuously develop and maintain the simulation software
and Geant4 in particular. We noted that Geant4 is not only used to perform simulations
of experiments, but it is also used as a de-facto standard when
developing other simulation tools (e.g., ML-based).
We identified several challenges facing the simulation community, 
including concerning discontinuities and dormancy of some of the physics developments, 
(Subsection~\ref{subsec:modelingNeeds} and Section~\ref{sec:experts})
and described several technical efforts (Section~\ref{sec:multiprong}) undertaken to address some 
of the needs. We provided an example to enable an estimation of the magnitude 
of the human effort needed to develop and support the main software simulation tools needed 
to successfully carry out the simulation tasks for the current and future US HEP experiments (Section~\ref{sec:experts}).

\section{Acknowledgments}
\label{sec:ack}
This work was supported in part by the Fermi National Accelerator
Laboratory, managed and operated by Fermi Research Alliance, LLC under
Contract No. DE-AC02-07CH11359 with the U.S. Department of Energy (Fermilab publication number for this paper is FERMILAB-FN-1151-ND-PPD-SCD) and by the U.S. Department of Energy’s Office of Science,
Office of High Energy Physics, of the US Department of Energy under Contract No. KA2102021.

\bibliography{refs}

\end{document}